\documentclass[12pt,a4paper]{article}
\usepackage{amsmath}
\usepackage{amsfonts}
\usepackage{amssymb}
\usepackage{color}
\usepackage{epsfig, euscript, graphics}

\author{Andrei Khrennikov}
\title{Unconditional quantum correlations do not violate Bell's inequality}

\begin{document}

\maketitle
\abstract{In this paper I demonstrate that the quantum correlations of polarization (or spin) observables used in Bell's argument 
against local realism have to be interpreted as {\it conditional} quantum correlations. By taking into account additional sources of randomness
in Bell's type experiments, i.e., supplementary to source randomness, I calculate (in the standard quantum formalism) the complete quantum correlations. The main message 
of the quantum theory of measurement (due to von Neumann) is that complete correlations can be essentially smaller than the conditional ones. Additional sources of randomness
diminish correlations. One can say another way around: transition from unconditional correlations to conditional can increase them essentially. This is true for both classical 
and quantum probability. The final remark is that classical conditional correlations do not satisfy Bell's inequality. Thus we met the following {\it 
conditional probability dilemma}: either to use the conditional quantum probabilities, as was done by Bell and others, or complete quantum correlations. However, in the first case
the corresponding classical conditional correlations need not satisfy Bell's inequality and in the second case the complete quantum correlations satisfy  Bell's inequality. Thus 
in neither case we have a problem of mismatching of classical and quantum correlations. It seems that the whole structure of Bell's argument was based on unacceptable identification 
of conditional quantum correlations with unconditional classical correlations.}

\section{Introduction}

Bell's argument \cite{B0}, \cite{B} against the local realism played the great role in the quantum foundational Renaissance and tremendous development of quantum technologies, especially 
in quantum information. One of the most attractive sides of this argument was its simplicity, logical, mathematical, and experimental (in the latter case, at least from the viewpoint
of the experimental design).  
By analyzing the probabilistic structure of Bell's argument I immediately understood that, from the purely probabilistic viewpoint, the whole Bell story is about interrelation
between conditional and unconditional probabilities, see, e.g., the first edition of my monograph \cite{INT0}. The experimental probabilities of Bell's type 
have to  be  compared with {\it conditional classical probabilities} \cite{K}. However, one cannot derive Bell's inequality for conditional (classical) probabilities. I tried to 
clarify this
problem of interrelation between conditional and unconditional probabilities in a long series of works by using a variety of arguments, see, e.g., 
\cite{CONT}--\cite{KHRX} and references herein. 
Now it becomes clear that the main problem was in concentration on the classical probabilistic counterpart of the problem. In particular, there was developed a very general 
contextual probability theory \cite{CONT} in which  Bell's inequality is violated, as well as other basic laws of classical (Kolmogorov \cite{K}, 1933) probability theory, e.g., 
the law of total probability. Violation of latter expresses interference in the purely probabilistic terms. 
 
In this paper I proceed by applying solely {\it the standard quantum formalism of measurement theory}, due to von Neumann \cite{VN}.       

It will be shown that the quantum correlations of polarization (or spin) observables used in Bell's argument 
against local realism have to be interpreted as {\it conditional} quantum correlations. By taking into account additional sources of randomness
in Bell's type experiments, i.e., supplementary to source randomness, cf. \cite{t1, KHRX} I calculate (in the standard quantum formalism) the complete quantum correlations. The main message 
of the quantum theory of measurement (due to von Neumann) is that the complete correlations can be essentially smaller than the conditional ones. Additional sources of randomness
diminish correlations. One can say another way around: transition from unconditional correlations to conditional can increase them essentially. This is true for both classical 
and quantum probability. The final remark is that classical conditional correlations do not satisfy Bell's inequality. Thus we met the following {\it 
conditional probability dilemma}: either to use the conditional quantum probabilities, as was done by Bell \cite{B0}, \cite{B} and others, or the complete quantum correlations. However, in the first case
the corresponding classical conditional correlations need not satisfy Bell's inequality and in the second case the complete quantum correlations satisfy  Bell's inequality. Thus 
in neither case we confront the problem of mismatching of classical and quantum correlations. It seems that the whole structure of Bell's argument was based on unacceptable identification 
of conditional quantum correlations with unconditional classical ones.

Therefore Bell's argument cannot be considered as an argument against local realism. From our viewpoint, the main message of quantum violation of Bell's inequality is encoded 
in the {\it Tsirelson bound}:
\begin{equation}
\label{TS}
\rm{UB}_{\rm{QM}}= 2\sqrt{2},
\end{equation}
 The classical probability theory cannot explain why conditional probabilities of some physical model violate Bell's inequality 
precisely up to this probabilistic constant $\rm{UB}_{\rm{QM}}.$ The classical theory gives for conditional probabilities the upper bound  
\begin{equation}
\label{TS1}
\rm{UB}_{\rm{CL}}= 4.
\end{equation} 

\section{Interrelation of observations on a compound system and its subsystems}

\subsection{Averages}

Consider a compound system ${\cal S}=(S, S^\prime),$ where $S$ and $S^\prime$ have the state spaces $H$ and $K,$ respectively; thus ${\cal S}$ is represented in the state 
space ${\cal H}= H \otimes K.$ Let $A_j, j=1,...,k,$ be a group of observables on $S,$ they are represented by Hermitian operators acting in $H$ which are denoted by 
the same symbols. In general these observables are incompatible. Consider also an observable $G$ on $S^\prime,$ having the values $j=1,...,k;$ it is represented by a Hermitian operator in $K,$ 
$G=\sum_j j P_j,$ where $(P_j)$ is its spectral family consisting of mutually orthogonal projectors. 

For any state $\rho,$  a density operator in $H,$ we can define the averages of observables $A_j: M_j=\rm{Tr} \rho A_j,$ and,
for any state $\sigma,$  a density operator in $H,$  the averages of observables $P_j: g_j=\rm{Tr} \sigma P_j.$   

Now we consider the observables ${\cal A}_j$ on the compound system ${\cal S}$
\begin{equation}
\label{OB}
{\cal A}_j = A_j \otimes P_j.
\end{equation}
Consider a state $R$ of ${\cal S}.$ We can define the averages of the observables ${\cal A}_j$ for this state:
\begin{equation}
\label{OB1}
m_j= \rm{Tr} R {\cal A}_j.
\end{equation}
Let  the state of the compound system $S$ be factorisable, i.e., its subsystems are not entangled,  
\begin{equation}
\label{OB2}
R= \rho \otimes \sigma.
\end{equation}
Then 
\begin{equation}
\label{OB3}
m_j  = M_j g_j .
\end{equation}

Suppose now that by experimenting with the compound system ${\cal S}$ one ``forgot'' about the presence of the subsystem $S^\prime.$

\medskip

{\it This forgetfulness has an interesting probabilistic effect: it induces the increase of averages, from $m_j$ to $M_j$ 
with the scaling coefficient $k_k= 1/g_j.$ }

\medskip

Thus if one treats an experiment on the compound system ${\cal S}$ as an experiment on its proper subsystem $S,$ the 
averages and  probabilities can  
increase essentially. For example, let $\sigma= I /\rm{dim} \;K.$ Then $g_j=\rm{dim} \;P_j/ \rm{dim} \;K$ and $k_j=\rm{dim} \; K/ \rm{dim} P_j.$ We are, in fact, interested in the case
$\rm{dim} \;K=4$ and $\rm{dim}\; P_j=1,$ i.e., $k_j=4,$ the {\it four times increase of the magnitudes} of the  averages and  probabilities. 

\medskip

This increase of averages explains ``the mystery of violations of Bell's type inequalities and superstrong quantum correlations'' (of course, only 
for a reader who is ready for my argument). 

\subsection{Correlations}
\label{corr}

Now move to the case of quantum correlations. Let now $H=H_1 \otimes H_2,$ i.e., $S$ is by itself a compound system $S=(S_1, S_2),$ and let 
$K=K_1 \otimes K_2,$ i.e., $S^\prime$ is by itself a compound system $S^\prime=(S_1^\prime, S_2^\prime).$ For the state space $H_1,$ we consider a pair of observables
$A_0, A_1$ and, for the state space $H_2,$ a pair of observables $B_0,B_1$; for  $K_1,$ a pair of observables represented by 
orthogonal projectors $P_0, P_1$ and, for $K_2,$  a pair $Q_0,Q_1.$  
Finally, let $\rho$ and $\sigma$ be the states represented by density operators acting in $H=H_1\otimes H_2$ and $K=K_1\otimes K_2.$ 

In Bell's experimental scheme the observables in $H_i$ represent polarization measurements 
and the observables in $K_i$ represent measurements of outputs of random generators. The state $\rho$ is the state of a pair of entangled photons  $S=(S_1, S_2)$
and the state $\sigma$ is a separable state of the pair of random generators, where the state of each random generator can (but need not) be given by a classical 
statistical mixture of two possible outputs. Of course, our scheme works for observables and random generators with an arbitrary number of outputs. By restricting
the numbers of outputs to two we just try to keep closer to our concrete aim -- Bell's experimental scheme.    

We introduce correlations in $H$
\begin{equation}
\label{OB4}
C_{ij}=\rm{Tr} A_i \otimes B_j \rho = \rm{Tr} O_{ij} \; \rho,
\end{equation}
where $O_{ij}= A_i \otimes B_j,$  
and in $K$          
\begin{equation}
\label{OB5}
g_{km} =  \rm{Tr} P_k \otimes Q_m \sigma= \rm{Tr} O_{km}^\prime \; \sigma, 
\end{equation}
where $O_{km}^\prime= P_k \otimes Q_m.$

Now we consider the state space of the compound system ${\cal S}=(S,  S^\prime)$ given by 
${\cal H}= H\otimes K= H_1 \otimes H_2 \otimes K_1 \otimes K_2.$ 
For any state given by a density operator $R$ on ${\cal H},$ we can find the correlation of the observables given by the operators $O_{ij}$ and $O_{km}^\prime:$
\begin{equation}
\label{OB6}
c_{ij} = \rm{Tr} O_{ij} \otimes O_{km}^\prime R = \rm{Tr} A_i \otimes B_j \otimes P_k \otimes Q_m R.
\end{equation}
Suppose that states in $H$ and $K$ are not entangled, i.e., $R= \rho\otimes \sigma.$ Then  
\begin{equation}
\label{OB7}
c_{ij, km} =\rm{Tr} O_{ij} \rho \; \rm{Tr} O_{km}^\prime = C_{ij} \; g_{km}
\end{equation}

Suppose again that by experimenting with the compound system ${\cal S}$ one ``forgot'' about the presence of the subsystem $S^\prime.$
If the correlation $g_{km; \sigma} <1,$ then: 

\medskip

{\it This forgetfulness  induces the increase of correlations, from $c_{ij, km}$ to $C_{ij; \rho}.$}

\medskip

In this way one obtain ``superstrong nonclassical Bell correlations''.

\subsection{Towards proper quantum formalization of Bell's experiment}
\label{LL}

Consider the case of the two dimensional spaces $K_1$ and $K_2.$  The corresponding bases in $K_t, t=1, 2,$  
are $(\vert 0\rangle_t,\vert 1\rangle_t).$ To shorter notation, further we omit the index $t.$
Let $P_\alpha, Q_\alpha=\vert \alpha\rangle \langle \alpha\vert,  \alpha=0,1.$  Let the  states in $K_1$ and $K_2$ are neither entangled, i.e.,
$\sigma= \sigma_1 \otimes \sigma_2$ and each state $\sigma_i$ is the ``classical mixture'':
\begin{equation}
\label{OB8}
\sigma_1=p_0 \vert 0\rangle \langle 0\vert + p_1 \vert 1\rangle \langle 1\vert, \;
\sigma_2=q_0 \vert 0\rangle \langle 0\vert + q_1 \vert 1\rangle \langle 1\vert, 
\end{equation}
where $p_0+p_1=1$ and $q_0+ q_1=1$ and nonnegative. Then 
$$
g_{km}= p_k q_m <1.   
$$
In particular, if all probabilities are equal, we obtain that 
\begin{equation}
\label{OB9}
g_{km} =1/4.
\end{equation}
This imply 4-times increase of correlations as the result of ``missing'' the subsystem $S^\prime.$ 

In the Bell-type experiments, e.g., for the CHSH-inequality,  one operates with the linear combination of correlations
\begin{equation}
\label{OB10}
C=C_{00}+C_{01}+ C_{10} - C_{11} 
\end{equation}
It is convenient to represent $C$ as the average of a single observable represented as the operator 
\begin{equation}
\label{OB10}
\Gamma=O_{00}+O_{01}+ O_{10} - O_{11}= A_0 \otimes B_0 +  A_1 \otimes B_0 + A_0 \otimes B_1
- A_1 \otimes B_1.
\end{equation}
Thus
\begin{equation}
\label{OB10a}
C=\rm{Tr} \rho \Gamma. 
\end{equation}
However, one can proceed with this operator only by ignoring the second subsytem of the compound system ${\cal S}=(S, S^\prime).$ By considering measurement on
${\cal S}$ and by taking into account correlations of observables on $S^\prime$ we come to the representation of the corresponding modification of the correlation 
function $C$ in the following operator form:
\begin{equation}
\label{OB11}
\gamma=O_{00}\otimes O_{00}^\prime + O_{01}\otimes O_{01}^\prime + O_{10}\otimes O_{10}^\prime - O_{11}\otimes O_{11}^\prime.
\end{equation}
The corresponding correlation function is given by the average
\begin{equation}
\label{OB11a}
c=\rm{Tr} \rho\otimes \sigma \;  \gamma= c_{00} + c_{01}  + c_{10}  - c_{11}, 
\end{equation}
where 
\begin{equation}
\label{OB11at}
c_{ij}= C_{ij} g_{ij}
\end{equation}
In the case of equal probabilities $p_i,q_j,$ see (\ref{OB9}), we have:   
\begin{equation}
\label{OB11a}
c=C/4. 
\end{equation}
Thus by taking into account that the second system $S^\prime$ is also involved in the experiment we find that the ``Bell correlation'' function $C$ decreases four times.

In particular, for the EPR-Bell correlations the Tsirelson bound for $C,$ namely, $C_{\rm{max}}=2\sqrt{2},$ which is the bound for conditional quantum correlations,
 leads to the following bound for unconditional quantum correlations
\begin{equation}
\label{OB11a}
c=\frac{\sqrt{2}}{2} <2. 
\end{equation}

\section{Complete description of systems and  observables involved in Bell's type experiments}

As was already emphasized, we want to account all physical systems and  observables which are involved in   Bell's type experiments.
Our main point is that in the standard quantum mechanical presentations, e.g., in textbooks (but as well as in research papers)
people forget about important physical systems playing the crucial role in the experiments. These are the random generators which 
outputs  determine orientations of PBSs

We shall proceed with the CHSH inequality. For our purpose, it is useful to modify the modern experimental scheme. Typically one uses only two PBSs to realize 
four orientations, two at one side and two at another side. The needed two orientations for each of PBSs are produced with the help of two random generators 
$G_1$ and $G_2.$ Here $G_1=i, i=0,1,$ leads to the orientation $i$ and hence the observation of $A_i$ and $G_2=j, j=0,1,$ leads to the orientation $j$ and 
hence the observation of $B_j.$

The crucial point is that {\it the outputs of random generators} also have to be considered as the results of measurements.
Of  course, generation of the pseudo-random numbers by a computer program which is often explored in Bell's type experiments dimmed  the role 
of these observables. However, even in this case the values  $G_1=i$ and $G_2=j$ have to be determined, so they can be treated as the results of the measurements
of the reading type. If one uses physical devices giving random numbers, then the treatment of production of random numbers as measurements is straightforward.
  
This measurement viewpoint to outputs of random generators is better visible in the modified experimental design in which each setting is represented by its 
own PBS: the design with two PBSs at each side and each PBS is equipped with its own pair of detectors - in total 4 PBs and 8 detectors combined with two 
stations  $C_1$ and $C_2$ distributing signals from the source in accordance with the results ot ``random generators measurements''. 

Such an experimental design was, in particular, used by A. Aspect \cite{Aspect} (but with just one detector for each of four PBSs).

\begin{figure}[htpb]
\centering{}
\includegraphics[scale=1]{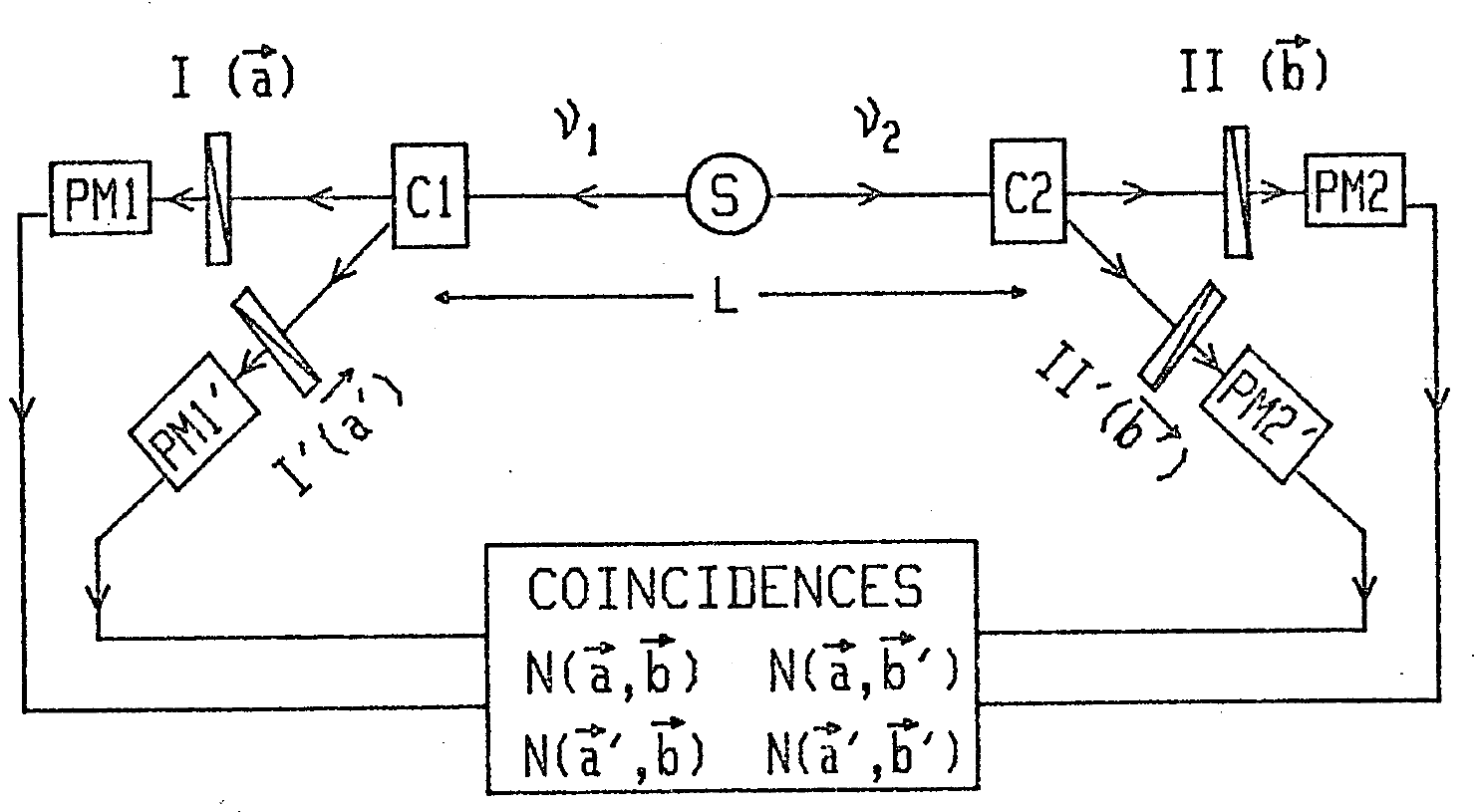} 
\caption{The scheme of the pioneer experiment of A. Aspect with four  beam splitters \cite{Aspect}.}
\label{fig1_asp} 
\end{figure}

We present the corresponding citation  
of  Aspect \cite{Aspect1}, see also \cite{Aspect},  section ``Difficulties of an ideal experiment'':

``We have done a step towards such an ideal experiment by using the modified scheme shown on Figure 15.
In that scheme, each (single-channel) polarizer is replaced by a setup involving a
switching device followed by two polarizers in two different orientations: $a$ and $a^\prime$ on side
I, $b$ and $b^\prime$ on side II. The optical switch $C1$ is able to rapidly redirect the incident light
either to the polarizer in orientation $a,$ or to the polarizer in orientation $a^\prime.$ This setup is
thus equivalent to a variable polarizer switched between the two orientations $a$ and $a^\prime.$ A
similar set up is implemented on the other side, and is equivalent to a polarizer switched
between the two orientations $b$ and $b^\prime.$ In our experiment, the distance $L$ between the two
switches was 13 m, and $L / c$ has a value of 43 ns. The switching of the light was effected by home built devices, based on the
acousto-optical interaction of the light with an ultrasonic standing wave in water. The
incidence angle (Bragg angle) and the acoustic power, were adjusted for a complete
switching between the 0th and 1st order of diffraction.''

So, there is no fundamental difference in proceedings with two or four PBSs. (Of course, we also can present our argument for experiments with only two 
PBSs, but here the measurement feature of random generation would be shadowed.)

\subsection{Taking into account random choice of settings}
\label{4_8}

As in \cite{KHRX}, we consider the following experimental design:

\begin{itemize}
\item[a).] There is a source of entangled photons. 

\item[b).] There are 4 PBSs and corresponding pairs of detectors for each PBS, totally 8 detectors. PBSs are
labeled   as $i= 1,2$ (at the left-hand side, LHS) and $j=1,2$ (at the right-hand side, RHS). 

\item[c).] Directly after source there are 2 distribution devices, one at LHS and one at RHS.
At each instance of time, $t=0, \tau, 2\tau, \ldots$ each device opens 
the port to only one (of two) optical fibers going to  the
corresponding two PBSs. For simplicity, 
These switches are controlled by two random generators $G_1$ (the left-hand side) and $G_2$ 
(the right hand side ) with probabilities of for the $i$-channel, $i=0,1,$ given by $p_i$ and $q_i,$ 
respectively ($p_1+p_2 =1, q_1 + q_2=1).$
\end{itemize}

Now we introduce the physical observables measured in this experiment.
 
1) ${\cal A}_i = \pm 1, i=0,1$ if  the $i$-th channel  (at LHS) is open and the corresponding (up or
down)  detector  fires; 

2) ${\cal A}_i= 0$ if the $i$-th channel (at LHS) is  blocked. 

In the same way we define the ``RHS-observables'' ${\cal B}_j=0, \pm 1,$
corresponding to PBSs $j=1,2$.

Thus unification of 4 incompatible experiments of the CHSH-test into a single experiment modifies the range of values
of polarization observables for each of 4 experiments; the new value, zero, is added to reflect the random choice of experimental 
settings. We emphasize that this value has no relation to the efficiency of detectors. In this model we assume that detectors have 
100\% efficiency. The observables take the value zero when the optical fibers going to the corresponding PBSs are blocked.

The measurement of the product of the observables ${\cal A}_i {\cal B}_j$ is represented by the quantum operator $O_{ij}\otimes O_{ij}^\prime,$
where $A_i$ and $B_j$ are polarization observables corresponding to pairs of angles $\theta_0, \theta_1$ and $\theta_0^\prime, \theta_1^\prime$
and $P_i, Q_j$ are one dimensional projectors which were defined in section \ref{LL}.  The state $\rho$ is, for example, one of the Bell states.

Then the complete correlations are represented as $c_{ij},$ see (\ref{OB11at}), and the corresponding Bell correlation function $c,$ see (\ref{OB11a}),
 does not exceeds the upper boundary $2.$ 
This is the upper boundary for unconditional classical correlations (obtained by Bell). Thus by taking into account randomness of selection of experimental settings 
we eliminate mismatching between classical and quantum correlations which was emphasized by Bell.

\section{Quantum conditional correlations}

Consider the same compound system ${\cal S}=(S, S^\prime)$ as in section \ref{corr}. We now remark that the joint measurement of the observables    
mathematically represented by the operators $O_{ij}\otimes I$ and $I \otimes O_{km}^\prime$ can be treated as their sequential measurement. The correlations 
are the same. Now let us consider another problem -- to find (again with the aid of the quantum theory of measurement) {\it conditional correlations} of the observables $A_i$ and $B_j$ -
conditioned to the fixed outputs of the measurements of the observables given by $P_i$ and $Q_j.$ 

First we measure (on the subsystem $S^\prime$ of ${\cal S})$ the observable $O_{ij}^\prime= P_i \otimes Q_j.$ In the quantum formalism this can be treated as measurement 
on ${\cal S}$ of the observable given by $I\otimes O_{ij}^\prime.$  If the initial state of ${\cal S}$ is $R,$ then by getting the result $(P_i=1,Q_j=1)$ we know that the initial 
state is transformed to the post-measurement state:
\begin{equation}
\label{CO1}
R \to R_{ij}= \frac{(I\otimes O_{ij}^\prime) R (I\otimes O_{ij}^\prime)}{\rm{Tr} (I\otimes O_{ij}^\prime) R (I\otimes O_{ij}^\prime)}.
\end{equation}
Now in accordance with the quantum formalism for conditional measurements we measure the observable $O_{ij}\otimes I.$ The corresponding (conditional) 
average is given by 
\begin{equation}
\label{CO2}
C_{ij\vert \rm{cond}}= \rm{Tr} R_{ij} (O_{ij}\otimes I).
\end{equation}
If $R= \rho \otimes \sigma,$ then $(I\otimes O_{ij}^\prime) R (I\otimes O_{ij}^\prime) = \rho \otimes O_{ij}^\prime \sigma O_{ij}^\prime.$ In particular, 
the denominator in (\ref{CO1}) is given by $\rm{Tr} O_{ij}^\prime \sigma O_{ij}^\prime.$ We also have:
\begin{equation}
\label{CO2}
C_{ij\vert \rm{cond}}= \frac{\rm{Tr} \rho O_{ij} \; \rm{Tr} O_{ij}^\prime \sigma O_{ij}^\prime}{\rm{Tr} O_{ij}^\prime \sigma O_{ij}^\prime}= \rm{Tr} \rho O_{ij}.
\end{equation}
Hence, the term $\rm{Tr} O_{ij}^\prime \sigma O_{ij}^\prime$ disturbing Bell's argument peacefully disappeared.

Thus, if one treats the correlations $C_{ij}$ in Bell's correlation function $C,$ see (\ref{OB10}), as the quantum {\it conditional correlations}, then the diminishing effect 
discussed in section \ref{LL}  disappears.

Now one might argue that in Bell's argument it is really possible to  treat $C_{ij}$ as conditional quantum correlations. However, there is a dangerous pitfall on this way 
of reasoning, namely, that  Bell's inequality was proven for {\it unconditional  classical correlation} \cite{B}, see also \cite{CONT} for detailed analysis. 

\medskip

Is it possible to prove Bell's inequality for conditional classical
correlations? 
  
\medskip

The answer is no. By using conditional correlations one can easily violate Bell's inequality, see \cite{KHRX} for details.

\section{Concluding remarks}

By using the standard quantum formalism we demonstrated that complete quantum correlations in Bell's type experiment do not violate Bell's inequality.
It seems that Bell and following him scientists used improper quantum mechanical description of such experiments. The conditional quantum correlations, where
conditioning is to the choice of fixed experimental settings (e.g., the orientations of PBSs), were compared with unconditioned classical correlations. 

From our analysis, it is clear that in Bell's framework there are two scientifically justified ways of proceeding:

\begin{itemize}
\item either with conditional quantum correlations (which are used in the literature on Bell's argument) and then compare them with classical
conditional correlations,
\item or with complete (unconditional) quantum correlations and then compare them with classical unconditional correlations. 
\end{itemize} 

In the first case, Bell's argument is collapsed, since classical
conditional correlations can violate Bell's inequality; in the second case, it is collapsed, since complete (unconditional) quantum correlations
satisfy Bell's inequality.   

We remark that a few authors used (implicitly) classical conditioning (with respect to some parameters of the Bell-type experiments) 
to create classical models violating Bell's inequality, cf. with  \cite{A}--\cite{KK3}. (This statement is not about the validity 
of concrete models, but about the essence of the method in the use.) I also think 
that conditioning on histories is the cornerstone of the interpretation of violation of Bell's inequality in the consistent histories approach \cite{Griffiths}.

From my viewpoint, the main message of Bell's considerations is encoded in the Tsirelson bound, see (\ref{TS}). The main problem is to find a physical explanation of the appearance 
this number in terms of classical conditional probability.


\begin{thebibliography}{400}

\bibitem{B0} Bell J S  1964 On the Einstein Podolsky Rosen Paradox, Physics 1, 3, 195–200

\bibitem{B} Bell J S  1987 {\it  Speakable and Unspeakable in Quantum Mechanics}
(Cambridge: Cambridge Univ. Press)


\bibitem{INT0} A.   Khrennikov, {\it Interpretations of Probability.} VSP Int. Sc. Publishers, Utrecht/Tokyo, 1999;
De Gruyter, Berlin, 2009,  second edition (completed).

\bibitem{K} Kolmogoroff A N 1933  {\it Grundbegriffe der Wahrscheinlichkeitsrechnung}
(Berlin: Springer Verlag); English translation: Kolmogorov A N 1956 
{\it Foundations of Theory of Probability} (New York: Chelsea Publishing Company)


\bibitem{CONT} A.   Khrennikov, {\it Contextual approach to quantum formalism,} Springer, Berlin-Heidelberg-New York, 2009.

\bibitem{KK} A. Khrennikov, Frequency analysis of the EPR-Bell
argumentation. {\it Foundations of Physics,} {\bf 32}, 1159-1174 (2002).

\bibitem{KBX1} Khrennikov A 2008 Bell-Boole inequality: Nonlocality or probabilistic incompatibility 
of random variables? {\it Entropy} {\bf 10} 19-32 

\bibitem{t1} Avis D, Fischer P, Hilbert A, and Khrennikov A 2009 Single, Complete, 
Probability Spaces Consistent With EPR-Bohm-Bell Experimental Data, 
{\it Foundations of Probability and Physics-5}
 vol 750 (Melville, NY: AIP)  pp  294-301.

\bibitem{KHRX} A. Khrennikov, CHSH Inequality: Quantum probabilities as classical conditional probabilities. Found. Phys. link.springer.com/article/10.1007/s10701-014-9851-8 (2015)

\bibitem{VN}  Von Neumann J  ]1932 {\it Mathematische Grundlagen der Quantenmechanik} (Berlin-Heidelberg-New York: Springer).
English translation: 1955 {\it Mathematical Foundations of Quantum Mechanics} (Princeton: Princeton Univ. Press)


\bibitem{Aspect} Aspect A 1983 {\it Three experimental tests of Bell inequalities by the measurement 
of polarization correlations between photons} (Orsay)

\bibitem{Aspect1} Aspect A  Bell's Theorem: The Naive View of an Experimentalist. arXiv:quant-ph/0402001.  


\bibitem{A} Accardi L 1970  The probabilistic roots of the quantum
mechanical paradoxes. In:  Diner S,  Fargue D, Lochak G and
Selleri F (eds) {\it The Wave--Particle Dualism. A Tribute to Louis
de Broglie on his 90th Birthday}, pp. 47--55 (Dordrecht: D. Reidel Publ.
Company).

\bibitem{16} Accardi L 2005 Some loopholes to save quantum nonlocality. {\it Foundations
of Probability and Physics-3}  vol 750 (Melville, NY: AIP)  pp. 1-20.

\bibitem{HPL5} K. Hess, H. De Raedt, and K. Michielsen, Hidden assumptions in the derivation of the Theorem of Bell. Phys. Scr. 2012, 014002 (2012). 

\bibitem{D60} K. De Raedt, K. Keimpema, H. De Raedt, K. Michielsen, and S.
Miyashita, A local realist model for correlations of the singlet state,
Euro. Phys. J. B 53, 139 – 142 (2006).

\bibitem{D6} H. De Raedt, F. Jin, and K. Michielsen, Data analysis of Einstein-Podolsky-Rosen-Bohm 
laboratory experiments, The Nature of Light: What are Photons? V, edited by 
C. Roychoudhuri, H. De Raedt, A.F. Kracklauer, Proc. of SPIE Vol. 8832, 88321N-88321N-11 (2013). 

\bibitem{KK2} M Kupczynski,  Bertrand's paradox and Bell's inequalities. Phys.Lett.A  121 , 205-207(1987)

\bibitem{KK3a} M Kupczynski.  Entanglement and Bell inequalities. J.Russ.Laser Research 26, 514-23(2005)

\bibitem{KK3} K. Hess, Einstein Was Right! Pan Stanford, Singapore (2014).

\bibitem{Griffiths}  Griffiths R B 2002 {\it  Consistent quantum theory} (Cambridge, U.K.: Cambridge University Press)

\end{thebibliography}
\end{document}